\pdfoutput=1
\documentclass[10pt,twocolumn]{article}
\usepackage[a4paper,textwidth=7in,top=0.5in,bottom=0.5in,columnsep=0.25in]{geometry}
\IfFileExists{newtxtext.sty}{\usepackage{newtxtext,newtxmath}}{\usepackage{mathptmx}}
\usepackage{graphicx,booktabs,amsmath,siunitx,microtype,tabularx,array}
\usepackage[numbers,sort&compress]{natbib}
\usepackage{placeins,caption,xcolor,stfloats}
\usepackage[hidelinks]{hyperref}
\makeatletter
\renewcommand\section{\@startsection{section}{1}{\z@}{-2.5ex plus -0.5ex minus -0.2ex}{1.2ex plus 0.2ex}{\normalfont\large\bfseries}}
\renewcommand\subsection{\@startsection{subsection}{2}{\z@}{-2ex plus -0.5ex minus -0.2ex}{0.8ex plus 0.2ex}{\normalfont\normalsize\bfseries}}
\makeatother
\newcommand{\uS}{\,\si{\micro\siemens}}
\newcommand{\pp}{~pp}
\newcommand{\xgate}{x_{\mathrm{gate}}}

\graphicspath{{figures/}}
\newcommand{\figwide}[3]{\begin{figure*}[!tbp]\centering\includegraphics{#1}\caption{#3}\label{#2}\end{figure*}}

\newcommand{\AuthorNames}{Haochen Chai, Qixu Zhu, Siyao Li, and Fangfang Jiang\textsuperscript{*}}
\newcommand{\AuthorAffiliations}{College of Medicine and Biological Information Engineering, Northeastern University, Shenyang, China}
\newcommand{\AuthorEmails}{\begin{tabular}{c}Haochen Chai: \texttt{chaihc@mails.neu.edu.cn}; Qixu Zhu: \texttt{2619921@dundee.ac.uk}\\Siyao Li: \texttt{2619987@dundee.ac.uk}; Fangfang Jiang: \texttt{jiangff@bmie.neu.edu.cn}\end{tabular}}
\newcommand{\AuthorORCIDs}{Haochen Chai: ORCID 0009-0007-4785-2912; Qixu Zhu: ORCID 0009-0000-3169-5325; Siyao Li: ORCID 0000-0004-0368-6968}
\newcommand{\CorrespondingAuthor}{Fangfang Jiang}
\newcommand{\CorrespondenceEmail}{jiangff@bmie.neu.edu.cn}
\newcommand{\CompetingInterests}{The authors declare no competing interests}
\newcommand{\AuthorContributions}{\textbf{Haochen Chai:} Conceptualization, Methodology, Software, Formal analysis, Project administration, Writing -- original draft. \textbf{Qixu Zhu:} Methodology, Writing -- review \& editing. \textbf{Siyao Li:} Visualization, Data curation, Formal analysis. \textbf{Fangfang Jiang:} Funding acquisition, Supervision}
\newcommand{\FundingDetails}{This work is supported by the Fundamental Research Funds for the Central Universities (N26GFZ002), and in part by the National Natural Science Foundation of China (No:  62427807)}

\title{Artifact removal improves electrodermal waveforms but not downstream classification in a virtual-reality balance task}
\author{\AuthorNames\\\small \AuthorAffiliations\\\small \AuthorEmails\\\small \AuthorORCIDs\\\small Correspondence: \CorrespondingAuthor; \texttt{\CorrespondenceEmail}}
\date{}

\begin{document}
\maketitle

\begin{abstract}
\normalsize
Artifact removal routinely precedes the classification of electrodermal activity (EDA), on the assumption that a cleaner signal supports a better decision. We tested this assumption in a virtual-reality (VR) balance-disturbance task. A residual gating network was trained on a benchmark with expert-corrected EDA, frozen, and applied to VR recordings, where raw and gated signals were classified by five published time-series methods under identical leave-one-participant-out evaluation. On the benchmark the gate detected artifacts well (median record AUROC 0.94) and reduced error inside artifact regions by 17.8\%. In the VR task it did not improve classification. Changes in balanced accuracy ranged from $-1.35$ to $+0.93$ percentage points, no classifier improved and two lost accuracy, and all five were equivalent to raw input within $\pm3.32$ points. The benefit was lost between waveform and decision. The correction that lowered waveform error also reduced skin conductance response detection in all 43 benchmark records. Processing left 92.8\% of predictions unchanged, and the predictions it did change were corrected and corrupted at similar rates. The VR recordings also carried little contamination (an estimated 4.6\% of samples), and even perfect localization of deliberately injected artifacts recovered only 3.3 points in the most sensitive classifier. A pooled association between artifact level and accuracy (11.3 points) disappeared within participants (0.1 points), showing how differences between people can make cleaning look useful. Preprocessing should be judged by the decision it supports, against an unprocessed arm.
\end{abstract}
\noindent\textbf{Keywords:} electrodermal activity; artifact removal; virtual reality; equivalence testing; preprocessing evaluation; negative result

\section{Introduction}
Physiological sensing lets an immersive virtual-reality (VR) system follow how its user is responding without interrupting the interaction. Electrodermal activity (EDA) suits this role. It is inexpensive to record with wearable sensors and tracks sympathetic arousal through tonic skin conductance and transient skin conductance responses (SCRs) \citep{tronstad2022eda,boucsein2012recommendations,posadaquintero2020pipeline}. It is also easily disturbed by movement, because pressure and shifts at the electrode produce deflections that resemble responses \citep{bari2026artifacts,taylor2015edaexplorer}. Applied pipelines therefore remove artifacts before inference, and the step is seldom questioned. We ask whether it improves the decision the system is built to make.

There is good reason to doubt that it does. A balance disturbance changes movement and physiology at the same moment, so a correction that removes motion-related variation can also remove information that separates one user state from another. More fundamentally, an artifact handler is trained and scored on how closely it reproduces a clean waveform, whereas a classifier relies on whatever features separate its classes, and nothing guarantees that the two objectives agree.

Processing choices are known to shape EDA results. Multiverse analyses of SCR quantification give different effect estimates under different pipelines \citep{kuhn2022manyverse,sjouwerman2022multiverse}, filtering changes sensitivity \citep{privratsky2020filtering}, and the choice of model inversion changes predictive validity \citep{bach2014comparison,staib2015optimising}. Studies that compare processed with unprocessed signals on the same task are few and point in different directions, one reporting a benefit \citep{delity2026associations} and another a small loss \citep{rah2026vr}. In EEG, automated cleaning has rarely increased condition effects \citep{delorme2023eeg}, and component-based artifact rejection did not improve deep-network decoding in three brain--computer interface tasks \citep{kang2024ica}. In applied EDA work the comparison is usually absent. Of 13 studies in a documented sample that removed EDA artifacts before a downstream task, 2 compared processed and unprocessed signals with the same model and protocol (section~\ref{sec:audit}); in the others, the benefit of cleaning was taken for granted.

We tested artifact removal against the task it is meant to serve. A continuous residual gate was trained on a public benchmark with expert-corrected EDA \citep{edabe2023,llanesjurado2023lstmcnn}, frozen, and transferred to a VR balance-disturbance corpus \citep{ribeiro2025balance}, where raw and gated signals were classified by five published time-series methods under identical participant-disjoint folds. The gate performed well on the benchmark and gave no benefit in VR. Because the benchmark supplies expert references and the VR corpus supplies task labels, we could follow the expected benefit from contamination to decision and show where it was lost: in event preservation, in the direction of feature changes, in decisions that did not change, and in the small amount of contamination there was to remove. We also show that a pooled association between artifact level and accuracy, the kind of evidence often used to justify cleaning, reflects differences between participants rather than a benefit of cleaning.

\section{Methods}
\subsection{Data}
Two public corpora were used, each for what it can support (figure~\ref{fig:design}, table~\ref{tab:corpora}). EDABE~v2 \citep{edabe2023,llanesjurado2023lstmcnn} contains 43 EDA records sampled at 128~Hz, each with an expert-corrected reference waveform and a point-wise artifact mask. Annotated artifacts cover 10.75\% of all samples, and a fixed split assigns 33 records to training and 10 to testing. This corpus supports supervised training and a direct check of detection, reconstruction and event preservation.

The VR balance-disturbance corpus \citep{ribeiro2025balance} provides the task labels. After removal of one duplicated file (P11/P12), it contains 11 participants. Recordings include Xsens motion capture, EMG and Shimmer EDA; we used EDA at 58.8~Hz.

\begin{table}[tbp]\centering\small
\caption{Roles of the two corpora. The benchmark supports expert-referenced evaluation of the artifact handler; the VR corpus supplies the task labels.}
\label{tab:corpora}
\begin{tabularx}{\columnwidth}{l>{\raggedright\arraybackslash}X>{\raggedright\arraybackslash}X}\toprule
Property & EDABE v2 (benchmark) & VR balance disturbance\\\midrule
Unit & 43 records & 11 participants\\
Sampling rate & 128 Hz & 58.8 Hz\\
Reference & Expert mask and clean waveform & None\\
Evaluation & Fixed split, 33 training / 10 test records & 11-fold leave-one-participant-out\\
Task windows & None & 984 windows of 720 samples\\
Artifact burden & 10.75\% annotated & 4.64\% estimated\\\bottomrule
\end{tabularx}
\end{table}

\figwide{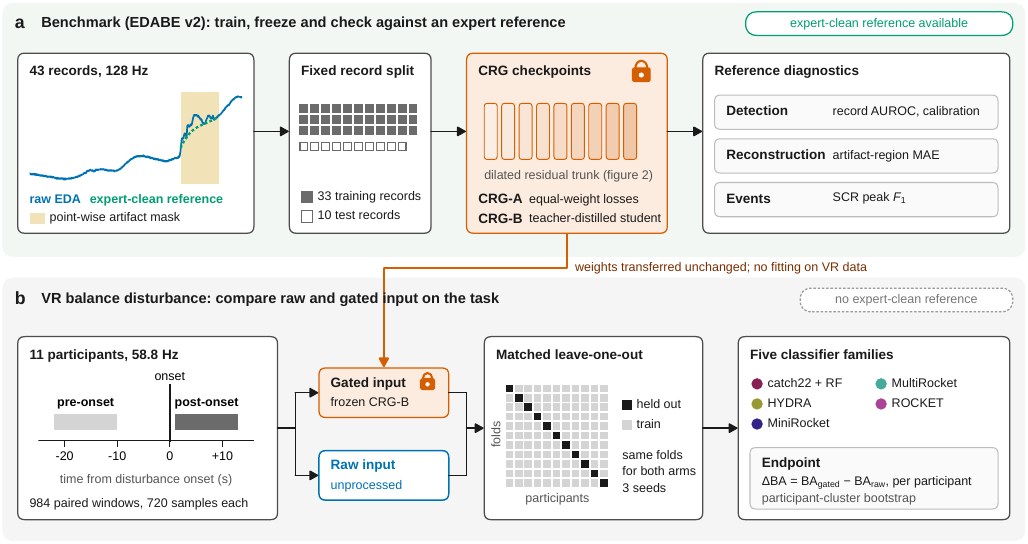}{fig:design}{Study design. (a) The benchmark (EDABE v2) provides expert-clean waveforms and point-wise artifact masks (thumbnail: record E01, raw trace with the expert reference inside the annotated span). Two checkpoints, CRG-A and CRG-B, were trained on the fixed split of 33 training and 10 test records and then frozen, and their detection, reconstruction and event preservation were scored against the reference. (b) The VR corpus provides labels for windows before and after each balance disturbance but no clean reference. The frozen CRG-B weights are applied without fitting on VR data, and raw and gated inputs pass through the same leave-one-participant-out folds (one held-out participant per fold, three seeds). The endpoint is the per-participant difference in balanced accuracy, with intervals from a participant-cluster bootstrap; family colours match figure~\ref{fig:task}.}

\subsection{Task}
Each disturbance onset defined a pair of windows: a positive window from 1 to 13~s after onset and a negative window from 22 to 10~s before it. A pair was discarded if either window crossed a trajectory boundary or overlapped another event, which kept the classes balanced within each participant. These rules were fixed before the labels were accessed. The final set contains 984 windows of 720 samples (12.24~s at 58.8~Hz). Because the positive window follows onset, the task separates the states before and after a disturbance rather than predicting a future loss of balance.

\subsection{Artifact handler}
\label{sec:handler}
The continuous residual gate (CRG; figure~\ref{fig:network}) corrects a trace by adding a learned residual weighted by a learned artifact probability:
\begin{equation}
\begin{aligned}
\tilde{x}_{\mathrm{gate}}(t)&=\max\{\tilde{x}(t)+p(t)\,r(t),\,0\},\\
p(t)&=\sigma(\ell(t)),\qquad r(t)=4\tanh(d(t)),
\end{aligned}
\label{eq:gate}
\end{equation}
where $\tilde{x}(t)=x(t)/s$ is the raw trace divided by a robust per-record scale,
\begin{equation}
s=\max\{P_{95}(x)-P_{5}(x),\;\SI{0.05}{\micro\siemens}\},
\label{eq:scale}
\end{equation}
and $\ell(t)$ and $d(t)$ are the outputs of a detection head and a residual head. Both heads share a temporal convolutional trunk with 48 channels and nine residual blocks with dilations from 1 to 256, which gives a receptive field of 4089 samples (about 32~s at 128~Hz). The corrected trace is multiplied by $s$ to return to microsiemens. Where $p(t)$ is near zero the trace passes unchanged, and the clamp keeps conductance non-negative.

Training combined a class-balanced cross-entropy for detection, an $L_1$ loss to the expert reference inside artifact regions, and an $L_1$ loss to the raw signal outside them; the last term discourages changes where none are needed. Two checkpoints were trained with this architecture on the same record split. CRG-A used the three losses with equal weights. CRG-B first trained a fold-local teacher for 12 epochs and then a student for 18 epochs whose objective added 0.15 times a binary cross-entropy to the detached teacher probabilities. All weights were frozen before any downstream analysis.

Each analysis uses the checkpoint suited to it. Detection, the waveform examples and the feature analysis use CRG-A. Benchmark reconstruction and event preservation both use the continuous-strength CRG-B student, so these two outcomes are compared within one checkpoint and one set of records. The VR task uses CRG-B, and the injection experiment applies CRG-A window by window.

Event preservation was measured on the benchmark. SCR peaks were detected on each signal after a second-order 0.05~Hz high-pass filter (prominence at least \SI{0.01}{\micro\siemens}, minimum separation 1~s) and matched one-to-one to peaks of the expert reference within $\pm1$~s. Peak $F_1$ summarizes this matching.

\figwide{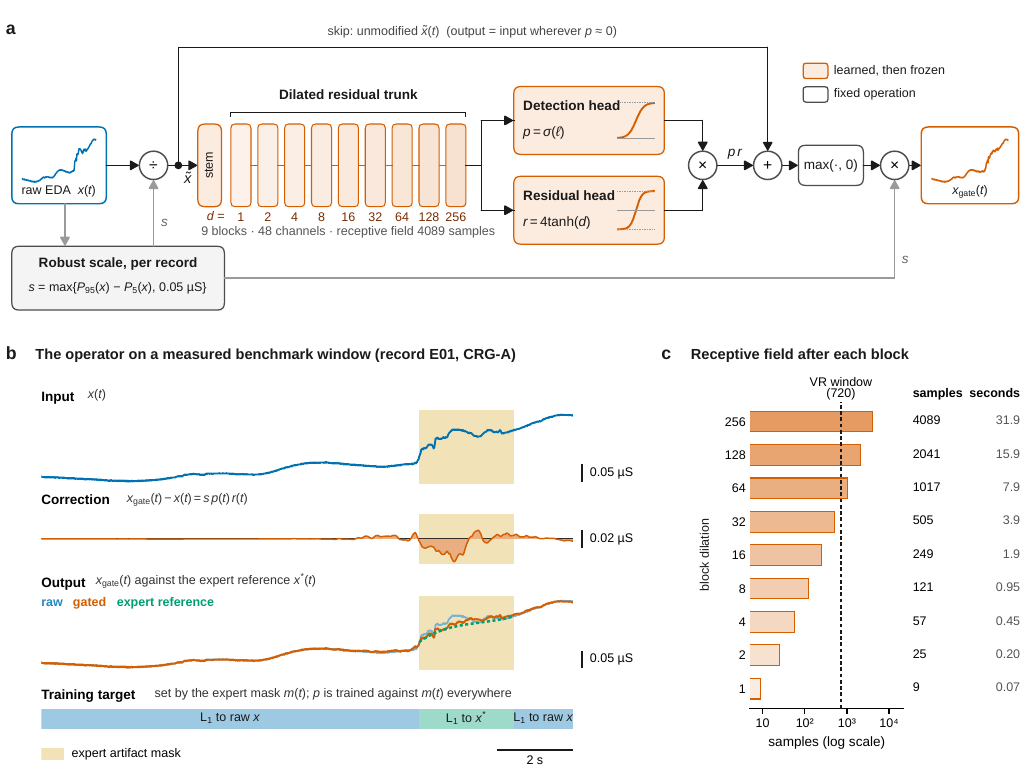}{fig:network}{The continuous residual gate. (a) The raw trace $x(t)$ is divided by a per-record robust scale $s$. A stem and nine dilated residual blocks (48 channels, dilation $d$ from 1 to 256) feed a detection head, whose output $p(t)=\sigma(\ell(t))$ lies between 0 and 1, and a residual head, whose output $r(t)=4\tanh(d(t))$ lies between $-4$ and 4. Their product is added to the unmodified normalized trace, clamped at zero and multiplied by $s$ (equations~\ref{eq:gate} and~\ref{eq:scale}), so the output equals the input wherever $p(t)$ is close to zero. Shaded modules are learned on the benchmark and then frozen; the other operations have no parameters. (b) The operator on a 14-s window of benchmark record E01 (CRG-A output, 128~Hz, unfiltered). The correction $\xgate-x=s\,p(t)\,r(t)$ is concentrated in the expert-annotated artifact span, where the output moves towards the expert reference $x^{*}(t)$. The bottom track shows the training targets that the mask defines: an $L_1$ loss to $x^{*}$ inside the span and to the raw signal outside it. The detection head is trained against the mask throughout, and CRG-B adds distillation from a fold-local teacher. (c) Receptive field after each block, from 9 samples after the first block to 4089 samples (31.9~s at 128~Hz) after the last. The dashed line marks the length of one VR window (720 samples).}

\subsection{Downstream evaluation}
Five published classifier families were used without modification: catch22 features with a random forest \citep{lubba2019catch22}, ROCKET \citep{dempster2020rocket}, MiniRocket \citep{dempster2021minirocket}, MultiRocket \citep{tan2022multirocket} and HYDRA \citep{dempster2023hydra}. They range from 22 interpretable features to large banks of random convolutional kernels, so a result that holds across them does not depend on one way of building features. The families were fixed before any treatment effect was seen.

Evaluation used 11-fold leave-one-participant-out cross-validation with seeds 1005, 2004 and 2027. Raw and gated arms used the same folds. Within each fold and seed, kernel transforms were fitted on training participants only and shared across arms, and each arm fitted its own scaler and classifier. The primary endpoint was the difference in balanced accuracy ($\Delta$BA), gated minus raw, in percentage points (pp). The decision analysis in section~\ref{sec:decisions} uses pooled accuracy instead, because pooled accuracy decomposes exactly into corrected and corrupted predictions.

\subsection{Statistics}
Participants were the unit of inference. Task-effect intervals come from a participant-cluster bootstrap with 20,000 resamples, with seeds averaged within each resample rather than treated as independent observations. Equivalence was tested with two one-sided tests (TOST) on seed-averaged participant differences. The margin of $\pm3.32$\pp\ is the gain that oracle repair achieved for the most sensitive classifier in the injection experiment (section~\ref{sec:injection}), and we also report the tightest margin at which all five families were equivalent. Bayes factors for a point null used a JZS prior with Cauchy scale $r=0.707$ against a two-sided alternative \citep{rouder2009bayest}. Design sensitivity is given as the minimum detectable effect for 80\% power at $\alpha=0.05$, $\mathrm{MDE}=2.80\times\mathrm{SE}$, where SE is the bootstrap standard error; the mean across families was 1.44\pp. Paired benchmark comparisons used Wilcoxon signed-rank tests across records.

For the analysis of participant composition, each VR window received a label-blind proxy for contamination: the mean absolute gated-minus-raw change divided by the record scale. Raw-input accuracy, averaged over models and seeds, was compared between low- and high-proxy windows after a median split over all windows and after a median split within each participant. The within-participant contrast weights participants equally, and bootstrap resamples kept repeated participants as distinct clusters.

\subsection{Controlled injection and contamination estimate}
To measure how much benefit repair can provide in this task, we added known artifacts to the VR windows. Artifact residuals, $a(t)=x_{\mathrm{raw}}(t)-x_{\mathrm{clean}}(t)$, were taken from expert-marked regions of the benchmark, resampled to 58.8~Hz and placed without reference to the labels:
\begin{equation}
x_{\mathrm{cont}}(t)=\max\{x_{\mathrm{vr}}(t)+s_{\mathrm{vr}}\,\hat{e}(t),\,0\},
\label{eq:injection}
\end{equation}
where $\hat{e}(t)$ is the placed residual in units of the record scale $s_{\mathrm{vr}}$. Dose is expressed as the effective perturbed fraction, the share of samples whose injected magnitude exceeds 0.01 of the record scale: 4.42\%, 8.94\% and 16.70\% for nominal doses of 5\%, 20\% and 40\%. At each dose, four arms were compared with the contaminated input. Oracle localization interpolated across injected samples using their known positions but not their values. The gate applied its learned correction. Matched noise added a phase-randomized surrogate with the same RMS and power spectrum as the gate's correction, which tests whether any change of that size helps. The clean-input arm used the native windows. Within each dose, fold and seed, MiniRocket fitted its transform on contaminated training data and shared it across arms; because the transform was refitted at each dose, dose comparisons describe the adaptive pipeline rather than the signal alone. The prespecified validity check was a loss for catch22 + RF at the highest dose relative to native input.

Natural contamination in the VR recordings was estimated with the transferred detector at a threshold of 0.930, where benchmark precision is 0.800 and recall 0.435. The flagged fraction was multiplied by precision and divided by recall. A second estimate held sensitivity and specificity fixed instead.

\subsection{Literature sample}
\label{sec:audit}
To see how often applied studies test preprocessing against the task, we searched PubMed, Europe PMC, Crossref and OpenAlex for studies published from 2019 to 2026 that recorded human EDA, handled artifacts and reported downstream inference \citep{bartolometomas2020musicelicited,jacobsen2021emotion,kong2021realtime,bhatkar2022statistical,hossain2022cognitive,yu2024statistical,aziz2023acute,amidei2023driver,gilmore2024human,nandipati2024mixedmethod,chen2026interpretable}. Of 363 candidates, 108 passed title screening and 80 were screened in full, leaving 13 eligible studies. A study counted as making the comparison if it evaluated processed and unprocessed signals with the same task, model and protocol; comparisons between processing methods, or signal-quality scores alone, did not count. The 13 studies form a documented, illustrative sample of current practice rather than an estimate of prevalence. One study was coded from its abstract alone and counted as not making the comparison; counting it the other way gives 3/13.

\section{Results}
\subsection{Gating did not improve classification}
Gating did not improve classification in any model family (figure~\ref{fig:task}a). Changes in balanced accuracy ranged from $-1.35$\pp\ for catch22 + RF to $+0.93$\pp\ for MiniRocket. No interval lay wholly above zero, and two lay wholly below it: catch22 + RF ($-1.35$\pp, 95\% CI $-2.38$ to $-0.25$) and HYDRA ($-1.00$\pp, $-1.98$ to $-0.05$). The most favourable upper bound was $+2.05$\pp, for MiniRocket, and in every family individual participants fell on both sides of zero.

The data also bound the effect. All five families were equivalent to raw input within $\pm3.32$\pp\ (all $p\le0.003$), and all five remained equivalent down to a margin of about $\pm2.37$\pp\ (figure~\ref{fig:task}b). With a mean design MDE of 1.44\pp, the comparison was sensitive to effects well below the size that would justify adding a preprocessing step. Bayes factors were inconclusive about an exact null (largest $\mathrm{BF}_{01}=2.88$, MultiRocket), and for catch22 + RF, whose estimate was negative, they leaned towards a nonzero effect ($\mathrm{BF}_{10}=2.08$). Gating therefore changed accuracy by less than two to three points, with no sign of benefit in any family. The rest of this section traces where the expected benefit was lost (table~\ref{tab:chain}).

\figwide{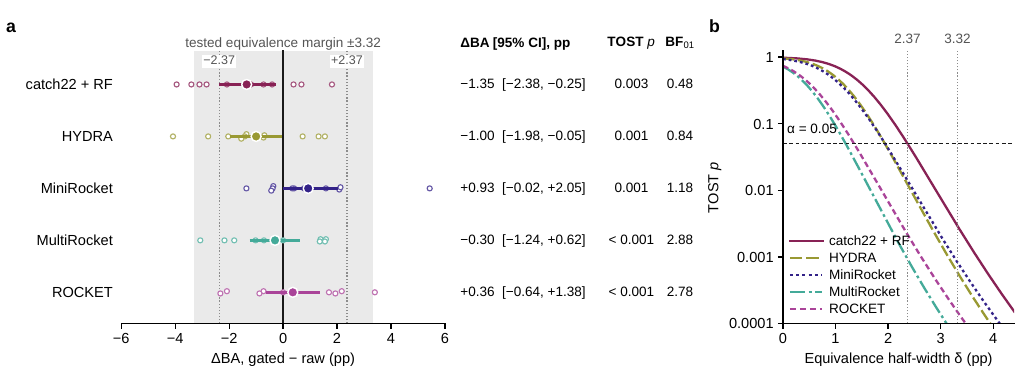}{fig:task}{Effect of gating on VR classification. (a) Gated-minus-raw balanced accuracy for each model family. Open circles are the 11 held-out participants, each averaged over seeds; filled circles and bars are the mean and its 95\% participant-cluster bootstrap interval. The shaded band is the $\pm3.32$\pp\ margin tested by TOST, and dotted lines mark the tightest common margin ($\pm2.37$\pp). $\mathrm{BF}_{01}$ uses a JZS prior ($r=0.707$) against a two-sided alternative. (b) TOST $p$ as a function of the equivalence half-width $\delta$, computed from the same participant differences (paired $t$, 10 degrees of freedom). A family is equivalent within $\pm\delta$ wherever its curve lies below $\alpha=0.05$.}

\begin{table*}[tbp]\centering\small
\caption{From contamination to decision: where the expected benefit was lost. Each row is a separate measurement on the checkpoint shown.}
\label{tab:chain}
\begin{tabularx}{\textwidth}{>{\raggedright\arraybackslash}p{0.13\textwidth}>{\raggedright\arraybackslash}p{0.2\textwidth}>{\raggedright\arraybackslash}X>{\raggedright\arraybackslash}p{0.2\textwidth}}\toprule
Stage & Measurement (checkpoint) & Result & Meaning\\\midrule
Contamination & Artifact share & 10.75\% annotated in the benchmark; 4.64\% estimated in VR & Less to remove in VR\\
Detection & Record AUROC (CRG-A) & Median 0.9409; 36/43 records above 0.90; median probability 3.13 times median artifact fraction & Good ranking, too much intervention\\
Reconstruction & Artifact-region MAE (CRG-B) & 0.1010 to 0.0830\uS\ ($-17.8$\%) & Waveforms improved\\
Events & SCR peak $F_1$ (CRG-B) & 0.8980 to 0.8674; lower in 43/43 records & Events degraded\\
Features & MiniRocket displacement (CRG-A) & No change needed in 57.4\% of windows, all changed; median cosine 0.363 & Change where none was needed, partial alignment where it was\\
Decisions & Prediction classes, VR (CRG-B) & 92.83\% unchanged; 3.46\% corrected; 3.71\% corrupted & Changes nearly cancel\\
Task & $\Delta$BA, VR (CRG-B) & $-1.35$ to $+0.93$\pp; equivalent within $\pm3.32$\pp & No benefit\\
Available benefit & Injection, catch22 + RF (CRG-A, window-wise) & Oracle $+3.32$\pp; extrapolated to native burden 1.05\pp & Little to recover\\\bottomrule
\end{tabularx}
\end{table*}

\subsection{The gate repaired waveforms on the benchmark}
Figure~\ref{fig:examples} shows two measured excerpts that illustrate how the correction can go wrong: it departs from the reference inside annotated spans at high artifact burden, and it intervenes where no artifact was annotated at low burden. Across all records, the detector ranked artifact samples well. The median record-level AUROC was 0.9409 (0.9477 in training records and 0.9263 in test records), and 36 of 43 records exceeded 0.90 (figure~\ref{fig:detector}a). Its probabilities, however, were too high. In every record the mean predicted probability exceeded the annotated artifact fraction (figure~\ref{fig:detector}c), and the median mean probability (0.209) was 3.13 times the median artifact fraction (0.067). The reliability curve lies well below the identity line over most of the probability range (figure~\ref{fig:detector}b). The detector thus ranks artifacts correctly but intervenes more often than the annotation warrants.

Reconstruction improved on average. Across the 43 records, artifact-region mean absolute error fell from 0.1010 to 0.0830\uS, a reduction of 17.8\% (paired Wilcoxon $p=1.2\times10^{-4}$; figure~\ref{fig:decoupling}a). The gain was concentrated in the records annotated by Expert~1, where error fell by 27.6\% and decreased in all 21 records; in Expert~2's records error was unchanged on average and fell in 10 of 22 (table~\ref{tab:subsets}).

\figwide{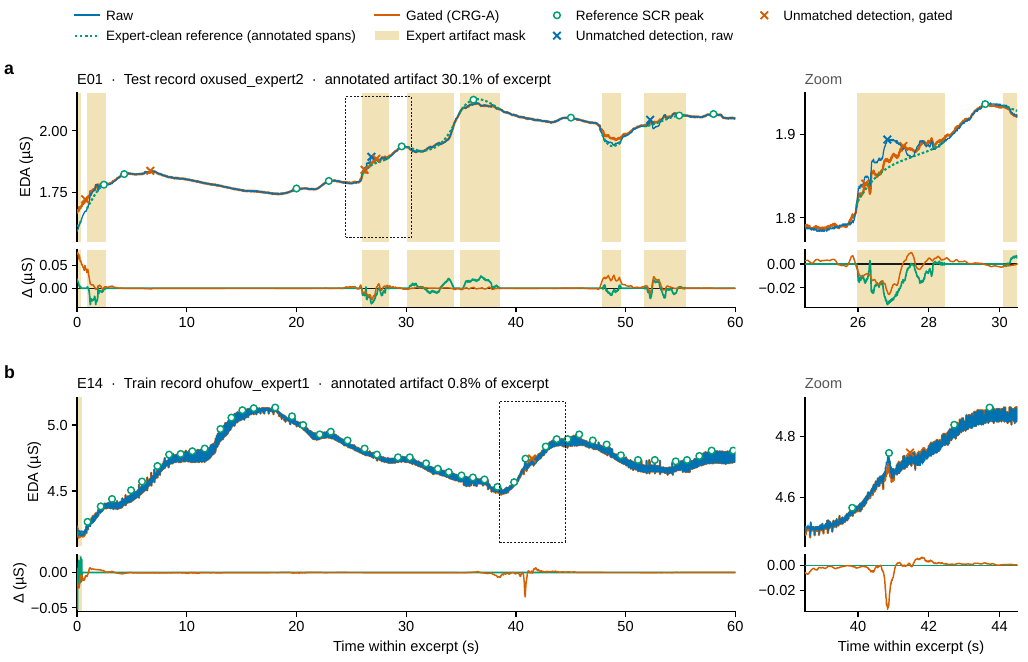}{fig:examples}{Two measured benchmark excerpts at 128~Hz, unfiltered, selected to illustrate failure modes rather than typical behaviour: gate failure at high artifact burden in E01 (test record) and over-intervention at low burden in E14 (training record). Upper panels show the raw trace (blue) and the CRG-A output (vermillion), which is drawn beneath raw so that it is visible only where the gate moved the signal. The expert-clean reference (green, dotted) is drawn within annotated spans; outside them it equals raw in E14 and differs from raw by less than \SI{0.0025}{\micro\siemens} in E01. Lower strips show reference minus raw (green) and gated minus raw (vermillion). Circles are reference SCR peaks, and crosses are detections on the raw or gated trace left unmatched by one-to-one matching within $\pm1$~s. Dashed boxes mark the windows enlarged at right.}

\figwide{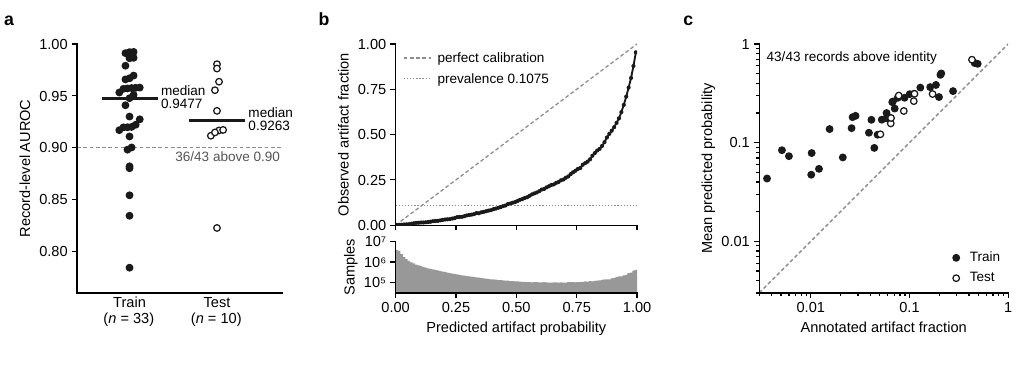}{fig:detector}{Detector behaviour on the benchmark (CRG-A; out-of-fold predictions for training records, test-ensemble predictions for test records). (a) Point-wise AUROC for each record; bars mark split medians, and 36 of 43 records exceed 0.90. (b) Observed artifact fraction in 100 fixed probability bins of width 0.01, with the number of samples per bin below on a log scale (34,312,745 samples in total). (c) Mean predicted probability against annotated artifact fraction for each record. All 43 records lie above the identity line.}

\subsection{Lower waveform error came with worse event detection}
The same CRG-B correction that lowered waveform error made SCR detection worse. Peak $F_1$ fell from 0.8980 to 0.8674 ($-3.05$\pp; $p=2.3\times10^{-13}$), and it fell in all 43 records, in both expert subsets and in both splits (figure~\ref{fig:decoupling}b, table~\ref{tab:subsets}). Figure~\ref{fig:decoupling}c shows the two outcomes together: 31 records gained waveform accuracy while losing event accuracy, and the other 12 lost both. The gated signals also contained more peaks. The total excess of detected over reference peaks rose from 11,108 to 12,989 ($+16.9$\%), and the median ratio of detected to reference peak counts rose from 1.078 to 1.086. A correction can bring a trace closer to the reference on average while adding small deflections that a peak detector reads as responses, so waveform error alone does not tell whether events are preserved.

\begin{table}[tbp]\centering\small
\caption{CRG-B reconstruction and event outcomes by expert subset and split. MAE is averaged across records; reduction $=100\times(1-\mathrm{MAE}_{\mathrm{gated}}/\mathrm{MAE}_{\mathrm{raw}})$. The last two columns count records. Expert and split subsets overlap, so rows are not additive. In the test split alone, $F_1$ fell by 2.50\pp\ ($p=0.002$) and MAE did not change significantly ($p=0.084$; paired Wilcoxon).}
\label{tab:subsets}
\setlength{\tabcolsep}{3.5pt}
\begin{tabular}{lrcrcc}\toprule
 & & MAE (\si{\micro\siemens}) & Red. & MAE & $F_1$\\
Subset & $n$ & raw $\to$ gated & (\%) & lower & lower\\\midrule
Expert 1 & 21 & 0.134 $\to$ 0.097 & 27.6 & 21/21 & 21/21\\
Expert 2 & 22 & 0.069 $\to$ 0.070 & $-0.2$ & 10/22 & 22/22\\
Training & 33 & 0.102 $\to$ 0.085 & 16.3 & 25/33 & 33/33\\
Test & 10 & 0.098 $\to$ 0.075 & 23.1 & 6/10 & 10/10\\\bottomrule
\end{tabular}
\end{table}

\figwide{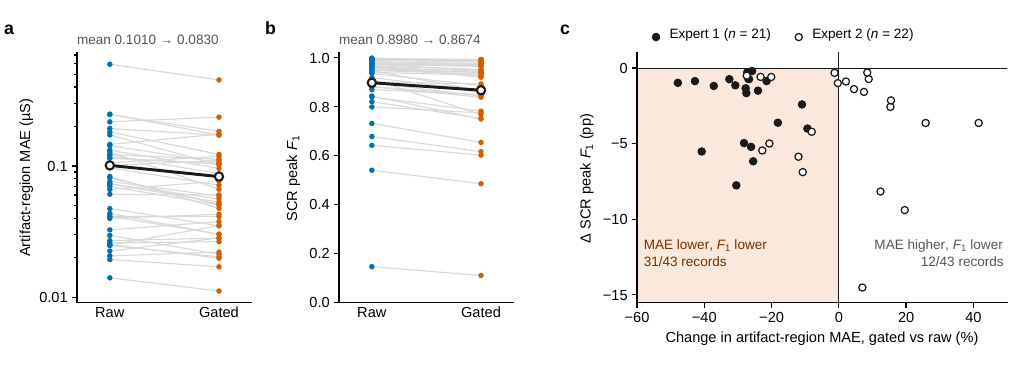}{fig:decoupling}{Waveform repair and event preservation on the 43 benchmark records, CRG-B continuous-strength student versus raw input. (a) Artifact-region mean absolute error to the expert-clean reference, on a log scale. (b) SCR peak $F_1$. Thin lines join each record and the thick line joins the means. (c) The same records as paired changes. Waveform error fell in 31 records and event $F_1$ fell in all 43. Filled circles are records annotated by Expert~1 and open circles those annotated by Expert~2.}

\subsection{Feature changes were widespread and weakly aligned}
Classifiers do not detect SCRs explicitly, so the next question is how the correction moved the features they use. For a frozen feature map $\phi$, we compared the displacement that the expert reference would require with the displacement the gate produced:
\begin{equation}
\Delta_{\mathrm{oracle}}=\phi(x_{\mathrm{clean}})-\phi(x_{\mathrm{raw}}),\qquad \Delta_{\mathrm{gate}}=\phi(\xgate)-\phi(x_{\mathrm{raw}}).
\label{eq:displacement}
\end{equation}
In 57.4\% of benchmark windows the reference required no change in MiniRocket features, yet the gate (CRG-A) changed the features of every one of these windows. Where change was needed, the gate moved features only partly in the right direction: the median cosine between the two displacements was 0.363 ($68.7^\circ$), and the gate's displacement was about half the required size (median ratio 0.507). The gate thus changed features where nothing needed changing and corrected them only partly where something did.

\subsection{Most decisions did not change}
\label{sec:decisions}
\begin{figure*}[!tbp]\centering
\includegraphics{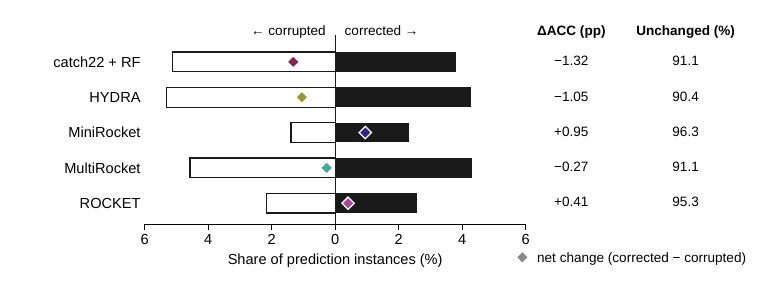}
\caption{Decision changes after gating for each model family, with 2,952 prediction instances per family (984 windows $\times$ 3 seeds). Bars to the right count instances that became correct, and bars to the left those that became incorrect. Diamonds mark the net change, which equals the pooled accuracy difference in equation~\ref{eq:identity}. The right-hand columns give that difference and the share of instances whose class did not change.}
\label{fig:decision}
\end{figure*}

Feature movement matters only if it changes a prediction. Each paired prediction either keeps its class, becomes correct or becomes incorrect, so the change in pooled accuracy is exactly
\begin{equation}
\Delta\mathrm{ACC}=\frac{N_{\mathrm{corrected}}-N_{\mathrm{corrupted}}}{N}.
\label{eq:identity}
\end{equation}
Averaged across families, 92.83\% of the 2,952 predictions per family (984 windows $\times$ 3 seeds) kept their class, 3.46\% became correct and 3.71\% became incorrect (figure~\ref{fig:decision}). The changes that did occur nearly cancelled. Across the five families, those that changed more predictions lost more accuracy (Spearman $\rho=-0.90$, exact $p=0.083$), as expected when the useful and harmful effects of a correction are of similar size.

\subsection{Little benefit was available to recover}
\label{sec:injection}
The last factor is how much contamination there was to remove. Corrected for benchmark precision and recall, the transferred detector implies an artifact burden of 4.64\% in the VR recordings; holding sensitivity and specificity fixed instead gives 2.88\%. Both estimates rest on the detector's benchmark performance, and both are well below the 10.75\% annotated in the benchmark.

The injection experiment measured what repair could achieve when contamination is known (figure~\ref{fig:injection}). For catch22 + RF, oracle localization recovered 3.32\pp\ relative to contaminated input, pooled over the three nonzero doses (95\% CI 1.86 to 4.76), whereas matched noise cost 2.11\pp\ ($-3.82$ to $-0.08$). The gate recovered 1.16\pp\ ($-1.13$ to 3.31), about a third of the oracle's benefit. For MiniRocket, oracle localization had no effect ($-0.29$\pp, $-1.81$ to 1.06). Repair helps a feature-based classifier when contamination is substantial and correctly located, and some classifiers are insensitive to it altogether.

At the highest dose, contamination lowered catch22 + RF accuracy by 4.13\pp\ relative to native input, with an interval that included zero ($-8.33$ to $+0.57$); the oracle and matched-noise contrasts show that the experiment registered both benefit and harm. Extrapolating the injection slopes to the estimated native burden gives expected benefits of 1.05\pp\ for catch22 + RF and 0.23\pp\ for MiniRocket. At the contamination present in this task, the benefit available even to perfect repair is small, below the mean design MDE of 1.44\pp.

\figwide{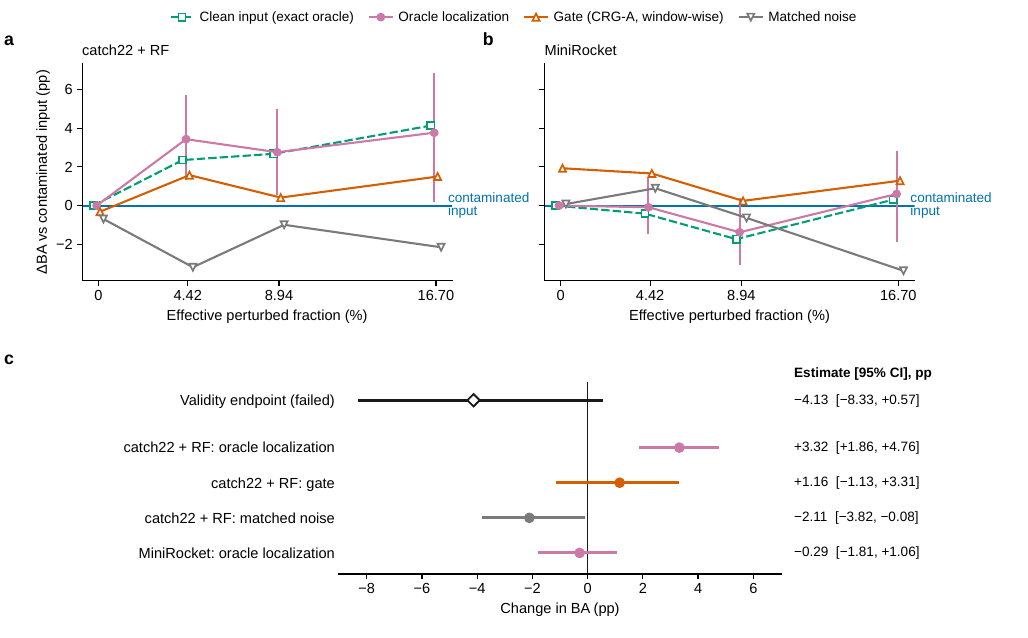}{fig:injection}{Controlled artifact injection. (a, b) Change in balanced accuracy relative to contaminated input at each effective perturbed fraction for catch22 + RF and MiniRocket; contaminated input defines the zero line. Points are means over 11 participants with seeds averaged, and arms are offset horizontally for legibility. Vertical bars are 95\% participant-cluster bootstrap intervals for the oracle-localization contrast. The clean-input arm reuses the native windows at every dose, so its curve shows the damage that perfect repair would undo; MiniRocket's values vary across doses only because its transform is refitted at each dose. (c) The prespecified validity check (catch22 + RF, highest dose versus native input) and pooled nonzero-dose contrasts against contaminated input.}

\subsection{Participant composition can make cleaning look useful}
\label{sec:confound}
A pooled association between artifact level and accuracy can make cleaning look necessary before any processed-versus-raw comparison is run. In our data, windows with low proxy values were classified 11.25\pp\ more accurately than windows with high values after a global median split (95\% CI 1.52 to 17.31). When the split was made within each participant, the difference was 0.07\pp\ ($-3.97$ to 4.10; figure~\ref{fig:confound}a). The pooled gap arose because high-proxy windows were concentrated in a few participants. The share of a participant's windows above the global median ranged from 9.2\% to 95.3\%, and three participants supplied 50.4\% of all high-proxy windows (figure~\ref{fig:confound}b). Participants who are harder to classify thus create an apparent link between artifacts and accuracy, and the pooled association describes who was measured rather than what cleaning does.

\figwide{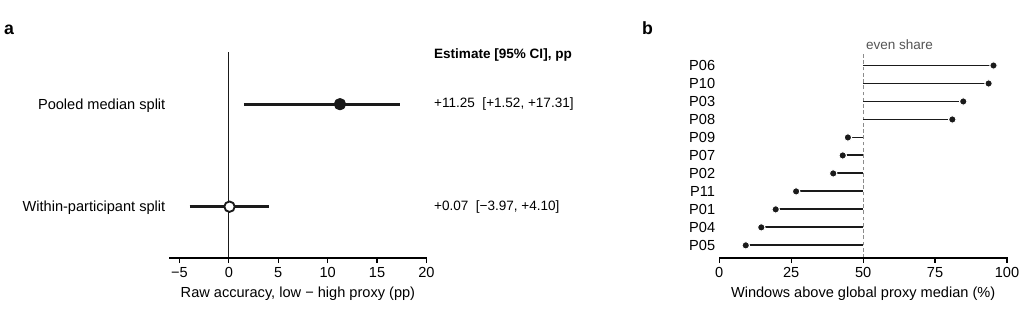}{fig:confound}{Participant composition behind the artifact--accuracy association. (a) Difference in raw-input accuracy between low- and high-proxy windows after a global median split and after a median split within each participant, with 95\% participant-cluster bootstrap intervals. (b) Share of each participant's windows above the global proxy median; without differences between participants every share would lie near 50\%. Three participants (P06, P10 and P03) supply 50.4\% of all high-proxy windows.}

\subsection{Sparser intervention did not preserve events}
The continuous gate changed almost every sample (99.8\%), which raises the question of whether a sparser rule would do better. A binary variant of the same student, which corrects only samples classified as artifact, changed 25.0\% of samples. It did not differ from the continuous gate in waveform error ($p=0.372$) or SCR $F_1$ ($p=0.071$). Both gates reduced $F_1$ relative to raw input, and the binary gate produced a higher ratio of detected to reference peaks ($p<10^{-4}$). Sparsity alone therefore does not preserve events. The more useful design target is the smallest correction that meets a fidelity requirement, checked against event and task outcomes rather than waveform error alone.

\section{Discussion}
\subsection{Why a better waveform did not give a better decision}
Artifact removal is usually justified by the quality of the cleaned waveform. In this study the waveform improved, yet the decision built on it did not, and the benchmark measurements show why. The correction that lowered waveform error added small deflections that degraded event detection in every record. It changed features in windows that needed no change and corrected them only partly in windows that did. Its effect on decisions was small and two-sided, so corrected and corrupted predictions nearly cancelled. The VR recordings, moreover, contained little contamination, so even perfect repair had little to recover. These effects act together and point to one conclusion: waveform fidelity is a poor guide to the value of preprocessing in a downstream task.

Repair quality in the VR recordings themselves could not be measured, because the VR corpus has no reference, and the move from 128~Hz benchmark records to 58.8~Hz VR recordings may have changed how the correction behaves. The practical conclusion holds either way. Anyone who applies a validated artifact handler to new data faces the same uncertainty, and only a processed-versus-raw comparison settles the question for their task.

\subsection{Relation to prior work}
Our findings join a growing body of work showing that preprocessing which improves a signal by its own criteria need not improve the analysis that follows. In EDA, analytic choices change SCR estimates \citep{kuhn2022manyverse,sjouwerman2022multiverse,delity2026associations}, and filtering and model inversion change sensitivity and predictive validity \citep{privratsky2020filtering,bach2014comparison,staib2015optimising}. In EEG, \citet{delorme2023eeg} found that, apart from high-pass filtering and bad-channel interpolation, automated corrections left the number of channels with significant condition effects unchanged or reduced it, and \citet{kang2024ica} found that component-based artifact rejection did not improve deep-network decoding in three brain--computer interface tasks. Reconstruction methods for EDA, from wavelet denoising \citep{chen2015wavelet} to learned models \citep{llanesjurado2023lstmcnn}, are evaluated against clean waveforms. That evaluation is necessary but, as our results show, not sufficient. This study adds a paired downstream comparison together with expert-referenced measurements of where the expected benefit is lost, from waveform to events, features and decisions, in a VR classification task.

\subsection{Implications for practice}
Three practices follow. Any EDA pipeline that removes artifacts should keep an unprocessed arm and compare the two on the intended decision, using identical windows, participant-disjoint folds, transforms and classifiers fitted on training participants only, and participant-level uncertainty; the outcome of this comparison should be reported whichever way it falls. A disappointing comparison should then be diagnosed through event preservation, the extent of intervention and the balance of corrected and corrupted decisions, not through waveform error alone. Finally, associations between artifact level and accuracy should be checked within participants before they are used to argue for cleaning. For the design of artifact handlers, our results favour the smallest correction that meets a fidelity requirement over corrections that change most of the signal. Systems meant to warn of a loss of balance before it happens will also need tests of lead time and prospective reliability, which a post-onset task does not address.

\subsection{Limitations}
Expert point-wise annotation exists for one corpus. Its two expert subsets do not overlap, so inter-rater agreement could not be assessed directly, although the subsets showed no significant differences in AUROC ($p=0.248$), artifact fraction ($p=0.202$) or change in SCR $F_1$ ($p=0.474$). With 11 participants, the downstream comparison bounds the effect rather than establishing its absence. The VR burden estimate assumes that detector precision and recall carry over from the benchmark, and it was computed on 120-s excerpts per participant. In the injection experiment the gate was applied to 720-sample windows, shorter than its receptive field of about 69.5~s at 58.8~Hz, so its absolute performance there differs from that of the full-trace gate, while comparisons across doses are unaffected. The decision analysis rests on five related model families. Finally, the study concerns one task, one artifact handler and five classifier families, and tasks with heavier contamination may gain more from repair.

\section{Conclusion}
Artifact removal improved electrodermal waveforms on a benchmark with expert references but did not improve classification of user state in a VR balance-disturbance task. The expected benefit was lost because waveform repair degraded event detection, changed features where no change was needed, left most decisions unchanged, and had little contamination to remove. Preprocessing should be judged by the decision it supports, with an unprocessed arm as the comparison and a within-participant check before artifact--accuracy associations are taken as evidence for cleaning.

\makeatletter
\renewcommand\section{\@startsection{section}{1}{\z@}{-1ex plus -0.2ex minus -0.1ex}{0.4ex plus 0.1ex}{\normalfont\normalsize\bfseries}}
\makeatother
\section*{Data and code availability}
Both corpora are public under a CC BY 4.0 licence: EDABE v2 (Mendeley Data, \url{https://doi.org/10.17632/w8fxrg4pv5.2}) and the VR balance-disturbance corpus (Zenodo, \url{https://doi.org/10.5281/zenodo.14013468}). The frozen analysis inputs, window-level predictions, analysis scripts and the code that regenerates every figure are available at \url{https://github.com/rtb-1005/FairEDA} and archived on Zenodo (\url{https://doi.org/10.5281/zenodo.23167890}).
\section*{Declaration of competing interest}
\CompetingInterests.
\section*{CRediT author contribution statement}
\AuthorContributions.
\section*{Funding}
\FundingDetails.
\section*{Ethics and data use}
This secondary analysis uses only publicly available data released under CC BY 4.0; ethical review was therefore not required.
\section*{Use of generative AI}
Generative AI tools (OpenAI Codex and Anthropic Claude) assisted with drafting, language editing and figure code. The authors reviewed and edited all output and take full responsibility for the content of this article.

\bibliographystyle{unsrtnat}
\bibliography{refs}

\begin{thebibliography}{35}
\providecommand{\natexlab}[1]{#1}
\providecommand{\url}[1]{\texttt{#1}}
\expandafter\ifx\csname urlstyle\endcsname\relax
  \providecommand{\doi}[1]{doi: #1}\else
  \providecommand{\doi}{doi: \begingroup \urlstyle{rm}\Url}\fi

\bibitem[Tronstad et~al.(2022)Tronstad, Amini, Bach, and
  Martinsen]{tronstad2022eda}
Christian Tronstad, Maryam Amini, Dominik~R. Bach, and Ørjan~G. Martinsen.
\newblock Current trends and opportunities in the methodology of electrodermal
  activity measurement.
\newblock \emph{Physiological Measurement}, 43\penalty0 (2):\penalty0 02TR01,
  2022.
\newblock \doi{10.1088/1361-6579/ac5007}.

\bibitem[Boucsein et~al.(2012)Boucsein, Fowles, Grimnes, Ben-Shakhar, Roth,
  Dawson, Filion, and {{Society for Psychophysiological Research Ad Hoc
  Committee on Electrodermal Measures}}]{boucsein2012recommendations}
Wolfram Boucsein, Don~C. Fowles, Sverre Grimnes, Gershon Ben-Shakhar, Walton~T.
  Roth, Michael~E. Dawson, Diane~L. Filion, and {{Society for
  Psychophysiological Research Ad Hoc Committee on Electrodermal Measures}}.
\newblock Publication recommendations for electrodermal measurements.
\newblock \emph{Psychophysiology}, 49\penalty0 (8):\penalty0 1017--1034, 2012.
\newblock \doi{10.1111/j.1469-8986.2012.01384.x}.

\bibitem[Posada-Quintero and Chon(2020)]{posadaquintero2020pipeline}
Hugo~F. Posada-Quintero and Ki~H. Chon.
\newblock Innovations in electrodermal activity data collection and signal
  processing: A systematic review.
\newblock \emph{Sensors}, 20\penalty0 (2):\penalty0 479, 2020.
\newblock \doi{10.3390/s20020479}.

\bibitem[Bari et~al.(2026)Bari, Yacoob~Aldosky, and
  Martinsen]{bari2026artifacts}
Dindar~S. Bari, Haval~Y. Yacoob~Aldosky, and Ørjan~G. Martinsen.
\newblock Electrodermal activity measurements: a review of artifacts.
\newblock \emph{Physiological Measurement}, 47\penalty0 (7):\penalty0 07TR01,
  2026.
\newblock \doi{10.1088/1361-6579/ae7bae}.

\bibitem[Taylor et~al.(2015)Taylor, Jaques, Chen, Fedor, Sano, and
  Picard]{taylor2015edaexplorer}
Sara Taylor, Natasha Jaques, Weixuan Chen, Szymon Fedor, Akane Sano, and
  Rosalind Picard.
\newblock Automatic identification of artifacts in electrodermal activity data.
\newblock In \emph{2015 37th Annual International Conference of the IEEE
  Engineering in Medicine and Biology Society ({EMBC})}, pages 1934--1937.
  IEEE, 2015.
\newblock \doi{10.1109/EMBC.2015.7318762}.

\bibitem[Kuhn et~al.(2022)Kuhn, Gerlicher, and Lonsdorf]{kuhn2022manyverse}
Manuel Kuhn, Anna M.~V. Gerlicher, and Tina~B. Lonsdorf.
\newblock Navigating the manyverse of skin conductance response quantification
  approaches: A direct comparison of trough-to-peak, baseline correction, and
  model-based approaches in ledalab and {PsPM}.
\newblock \emph{Psychophysiology}, 59\penalty0 (9):\penalty0 e14058, 2022.
\newblock \doi{10.1111/psyp.14058}.

\bibitem[Sjouwerman et~al.(2022)Sjouwerman, Illius, Kuhn, and
  Lonsdorf]{sjouwerman2022multiverse}
Rachel Sjouwerman, Sabrina Illius, Manuel Kuhn, and Tina~B. Lonsdorf.
\newblock A data multiverse analysis investigating non-model based {SCR}
  quantification approaches.
\newblock \emph{Psychophysiology}, 59\penalty0 (12):\penalty0 e14130, 2022.
\newblock \doi{10.1111/psyp.14130}.

\bibitem[Privratsky et~al.(2020)Privratsky, Bush, Bach, Hahn, and
  Cisler]{privratsky2020filtering}
Anthony~A. Privratsky, Keith~A. Bush, Dominik~R. Bach, Emily~M. Hahn, and
  Josh~M. Cisler.
\newblock Filtering and model-based analysis independently improve
  skin-conductance response measures in the {fMRI} environment: Validation in a
  sample of women with {PTSD}.
\newblock \emph{International Journal of Psychophysiology}, 158:\penalty0
  86--95, 2020.
\newblock \doi{10.1016/j.ijpsycho.2020.09.015}.

\bibitem[Bach(2014)]{bach2014comparison}
Dominik~R. Bach.
\newblock A head-to-head comparison of {SCRalyze} and ledalab, two model-based
  methods for skin conductance analysis.
\newblock \emph{Biological Psychology}, 103:\penalty0 63--68, 2014.
\newblock \doi{10.1016/j.biopsycho.2014.08.006}.

\bibitem[Staib et~al.(2015)Staib, Castegnetti, and Bach]{staib2015optimising}
Matthias Staib, Giuseppe Castegnetti, and Dominik~R. Bach.
\newblock Optimising a model-based approach to inferring fear learning from
  skin conductance responses.
\newblock \emph{Journal of Neuroscience Methods}, 255:\penalty0 131--138, 2015.
\newblock \doi{10.1016/j.jneumeth.2015.08.009}.

\bibitem[Delity et~al.(2026)Delity, Barhoum, Yu, Sherwood, Mischkowski, Necka,
  Lee, Dildine, Pereira, and Atlas]{delity2026associations}
Maya~Joshi Delity, Sabrina Barhoum, Qingbao Yu, Kai Sherwood, Dominik
  Mischkowski, Elizabeth~A. Necka, In-seon Lee, Troy~C. Dildine, Francisco
  Pereira, and Lauren~Y. Atlas.
\newblock A multiverse approach to heat-evoked skin conductance analysis:
  evaluating the influence of analytic pipeline on associations between skin
  conductance and pain.
\newblock \emph{Pain}, 167\penalty0 (8):\penalty0 1793--1809, 2026.
\newblock \doi{10.1097/j.pain.0000000000003996}.

\bibitem[Rah and Chen(2026)]{rah2026vr}
Audrey Rah and Yuhua Chen.
\newblock Electrodermal temperature-adjusted electrodermal activity ({EDA}) for
  stress detection in virtual reality.
\newblock \emph{Sensors}, 26\penalty0 (10):\penalty0 2983, 2026.
\newblock \doi{10.3390/s26102983}.

\bibitem[Delorme(2023)]{delorme2023eeg}
Arnaud Delorme.
\newblock {EEG} is better left alone.
\newblock \emph{Scientific Reports}, 13:\penalty0 2372, 2023.
\newblock \doi{10.1038/s41598-023-27528-0}.

\bibitem[Kang et~al.(2024)Kang, Chen, and Wallraven]{kang2024ica}
Taeho Kang, Yiyu Chen, and Christian Wallraven.
\newblock I see artifacts: {ICA}-based {EEG} artifact removal does not improve
  deep network decoding across three {BCI} tasks.
\newblock \emph{Journal of Neural Engineering}, 21\penalty0 (6):\penalty0
  066036, 2024.
\newblock \doi{10.1088/1741-2552/ad788e}.

\bibitem[Marín-Morales(2023)]{edabe2023}
Javier Marín-Morales.
\newblock Electrodermal activity artifact correction benchmark ({EDABE}).
\newblock Mendeley Data, 2023.
\newblock URL \url{https://data.mendeley.com/datasets/w8fxrg4pv5/2}.

\bibitem[Llanes-Jurado et~al.(2023)Llanes-Jurado, Carrasco-Ribelles, Alcañiz,
  Soria-Olivas, and Marín-Morales]{llanesjurado2023lstmcnn}
Jose Llanes-Jurado, Lucía~A. Carrasco-Ribelles, Mariano Alcañiz, Emilio
  Soria-Olivas, and Javier Marín-Morales.
\newblock Automatic artifact recognition and correction for electrodermal
  activity based on {LSTM-CNN} models.
\newblock \emph{Expert Systems with Applications}, 230:\penalty0 120581, 2023.
\newblock \doi{10.1016/j.eswa.2023.120581}.

\bibitem[Ribeiro et~al.(2025)Ribeiro, Veloso, Pires, and
  Santos]{ribeiro2025balance}
Nuno~Ferrete Ribeiro, Andr\'e Veloso, Henrique Pires, and Cristina~P. Santos.
\newblock Immersive ecological virtual environment for inducing balance
  disturbances.
\newblock \emph{IEEE Transactions on Visualization and Computer Graphics},
  31\penalty0 (12):\penalty0 10372--10383, 2025.
\newblock \doi{10.1109/TVCG.2025.3598464}.

\bibitem[Lubba et~al.(2019)Lubba, Sethi, Knaute, Schultz, Fulcher, and
  Jones]{lubba2019catch22}
Carl~H. Lubba, Sarab~S. Sethi, Philip Knaute, Simon~R. Schultz, Ben~D. Fulcher,
  and Nick~S. Jones.
\newblock {catch22}: {CAnonical} time-series {CHaracteristics}: Selected
  through highly comparative time-series analysis.
\newblock \emph{Data Mining and Knowledge Discovery}, 33\penalty0 (6):\penalty0
  1821--1852, 2019.
\newblock \doi{10.1007/s10618-019-00647-x}.

\bibitem[Dempster et~al.(2020)Dempster, Petitjean, and
  Webb]{dempster2020rocket}
Angus Dempster, François Petitjean, and Geoffrey~I. Webb.
\newblock {ROCKET}: exceptionally fast and accurate time series classification
  using random convolutional kernels.
\newblock \emph{Data Mining and Knowledge Discovery}, 34\penalty0 (5):\penalty0
  1454--1495, 2020.
\newblock \doi{10.1007/s10618-020-00701-z}.

\bibitem[Dempster et~al.(2021)Dempster, Schmidt, and
  Webb]{dempster2021minirocket}
Angus Dempster, Daniel~F. Schmidt, and Geoffrey~I. Webb.
\newblock {MiniRocket}: A very fast (almost) deterministic transform for time
  series classification.
\newblock In \emph{Proceedings of the 27th ACM SIGKDD Conference on Knowledge
  Discovery \& Data Mining}, KDD '21, pages 248--257. ACM, 2021.
\newblock \doi{10.1145/3447548.3467231}.

\bibitem[Tan et~al.(2022)Tan, Dempster, Bergmeir, and Webb]{tan2022multirocket}
Chang~Wei Tan, Angus Dempster, Christoph Bergmeir, and Geoffrey~I. Webb.
\newblock {MultiRocket}: multiple pooling operators and transformations for
  fast and effective time series classification.
\newblock \emph{Data Mining and Knowledge Discovery}, 36\penalty0 (5):\penalty0
  1623--1646, 2022.
\newblock \doi{10.1007/s10618-022-00844-1}.

\bibitem[Dempster et~al.(2023)Dempster, Schmidt, and Webb]{dempster2023hydra}
Angus Dempster, Daniel~F. Schmidt, and Geoffrey~I. Webb.
\newblock {HYDRA}: competing convolutional kernels for fast and accurate time
  series classification.
\newblock \emph{Data Mining and Knowledge Discovery}, 37\penalty0 (5):\penalty0
  1779--1805, 2023.
\newblock \doi{10.1007/s10618-023-00939-3}.

\bibitem[Rouder et~al.(2009)Rouder, Speckman, Sun, Morey, and
  Iverson]{rouder2009bayest}
Jeffrey~N. Rouder, Paul~L. Speckman, Dongchu Sun, Richard~D. Morey, and
  Geoffrey Iverson.
\newblock Bayesian $t$ tests for accepting and rejecting the null hypothesis.
\newblock \emph{Psychonomic Bulletin \& Review}, 16\penalty0 (2):\penalty0
  225--237, 2009.
\newblock \doi{10.3758/PBR.16.2.225}.

\bibitem[Bartolomé-Tomás et~al.(2020)Bartolomé-Tomás, Sánchez-Reolid,
  Fernández-Sotos, Latorre, and
  Fernández-Caballero]{bartolometomas2020musicelicited}
Almudena Bartolomé-Tomás, Roberto Sánchez-Reolid, Alicia Fernández-Sotos,
  José~Miguel Latorre, and Antonio Fernández-Caballero.
\newblock Arousal detection in elderly people from electrodermal activity using
  musical stimuli.
\newblock \emph{Sensors}, 20\penalty0 (17):\penalty0 4788, 2020.
\newblock \doi{10.3390/s20174788}.

\bibitem[Jacobsen et~al.(2021)Jacobsen, Hafli, Tronstad, and
  Martinsen]{jacobsen2021emotion}
Fredrik~A. Jacobsen, Ellen~W. Hafli, Christian Tronstad, and Ørjan~G.
  Martinsen.
\newblock Classification of emotions based on electrodermal activity and
  transfer learning - a pilot study.
\newblock \emph{Journal of Electrical Bioimpedance}, 12\penalty0 (1):\penalty0
  178--183, 2021.
\newblock \doi{10.2478/joeb-2021-0021}.

\bibitem[Kong et~al.(2021)Kong, Posada-Quintero, and Chon]{kong2021realtime}
Youngsun Kong, Hugo~F. Posada-Quintero, and Ki~H. Chon.
\newblock Real-time high-level acute pain detection using a smartphone and a
  wrist-worn electrodermal activity sensor.
\newblock \emph{Sensors}, 21\penalty0 (12):\penalty0 3956, 2021.
\newblock \doi{10.3390/s21123956}.

\bibitem[Bhatkar et~al.(2022)Bhatkar, Picard, and
  Staahl]{bhatkar2022statistical}
Viprali Bhatkar, Rosalind Picard, and Camilla Staahl.
\newblock Combining electrodermal activity with the peak-pain time to quantify
  three temporal regions of pain experience.
\newblock \emph{Frontiers in Pain Research}, 3, 2022.
\newblock \doi{10.3389/fpain.2022.764128}.

\bibitem[Hossain et~al.(2022)Hossain, Kong, Posada-Quintero, and
  Chon]{hossain2022cognitive}
Md-Billal Hossain, Youngsun Kong, Hugo~F. Posada-Quintero, and Ki~H. Chon.
\newblock Comparison of electrodermal activity from multiple body locations
  based on standard {EDA} indices’ quality and robustness against motion
  artifact.
\newblock \emph{Sensors}, 22\penalty0 (9):\penalty0 3177, 2022.
\newblock \doi{10.3390/s22093177}.

\bibitem[Yu et~al.(2024)Yu, Xu, Xiao, Xu, and Ming]{yu2024statistical}
Haiqing Yu, Minpeng Xu, Xiaolin Xiao, Fangzhou Xu, and Dong Ming.
\newblock Detection of dynamic changes of electrodermal activity to predict the
  classroom performance of college students.
\newblock \emph{Cognitive Neurodynamics}, 18\penalty0 (1):\penalty0 173--184,
  2024.
\newblock \doi{10.1007/s11571-023-09930-6}.

\bibitem[Aziz et~al.(2023)Aziz, Khan, Hirachan, Chetty, Goecke, and
  Fernandez-Rojas]{aziz2023acute}
Sumair Aziz, Muhammad~Umar Khan, Niraj Hirachan, Girija Chetty, Roland Goecke,
  and Raul Fernandez-Rojas.
\newblock "where does it hurt?": Exploring {EDA} signals to detect and localise
  acute pain.
\newblock In \emph{2023 45th Annual International Conference of the IEEE
  Engineering in Medicine \& Biology Society (EMBC)}, pages 1--5, 2023.
\newblock \doi{10.1109/embc40787.2023.10341157}.

\bibitem[Amidei et~al.(2023)Amidei, Spinsante, Iadarola, Benatti, Tramarin,
  Pavan, and Rovati]{amidei2023driver}
Andrea Amidei, Susanna Spinsante, Grazia Iadarola, Simone Benatti, Federico
  Tramarin, Paolo Pavan, and Luigi Rovati.
\newblock Driver drowsiness detection: A machine learning approach on skin
  conductance.
\newblock \emph{Sensors}, 23\penalty0 (8):\penalty0 4004, 2023.
\newblock \doi{10.3390/s23084004}.

\bibitem[Gilmore and Nasseri(2024)]{gilmore2024human}
Justin Gilmore and Mona Nasseri.
\newblock Human activity recognition algorithm with physiological and inertial
  signals fusion: Photoplethysmography, electrodermal activity, and
  accelerometry.
\newblock \emph{Sensors}, 24\penalty0 (10):\penalty0 3005, 2024.
\newblock \doi{10.3390/s24103005}.

\bibitem[Nandipati et~al.(2024)Nandipati, Pal, and
  Mitra]{nandipati2024mixedmethod}
Kishore~Kumar Nandipati, Sonika Pal, and Ritayan Mitra.
\newblock Electrodermal activity ({EDA}) morphologies and prediction of
  engagement with simple moving average crossover: A mixed-method study.
\newblock \emph{Sensors}, 24\penalty0 (14):\penalty0 4565, 2024.
\newblock \doi{10.3390/s24144565}.

\bibitem[Chen et~al.(2026)Chen, Lin, Zhou, and He]{chen2026interpretable}
Zhibiao Chen, Chang Lin, Shiqin Zhou, and Xiayun He.
\newblock Physiological stress in elderly residents of densely populated urban
  villages: A skin conductance study with interpretable machine learning
  modeling.
\newblock \emph{Buildings}, 16\penalty0 (2):\penalty0 248, 2026.
\newblock \doi{10.3390/buildings16020248}.

\bibitem[Chen et~al.(2015)Chen, Jaques, Taylor, Sano, Fedor, and
  Picard]{chen2015wavelet}
Weixuan Chen, Natasha Jaques, Sara Taylor, Akane Sano, Szymon Fedor, and
  Rosalind~W. Picard.
\newblock Wavelet-based motion artifact removal for electrodermal activity.
\newblock In \emph{2015 37th Annual International Conference of the IEEE
  Engineering in Medicine and Biology Society ({EMBC})}, pages 6223--6226.
  IEEE, 2015.
\newblock \doi{10.1109/EMBC.2015.7319814}.

\end{thebibliography}
\end{document}